\documentstyle[aps,pre,epsf,epsfig]{revtex}
\begin{document}

% uncomment for twocolumn
\twocolumn[
\hsize\textwidth\columnwidth\hsize\csname @twocolumnfalse\endcsname

\title{
Interface Motion and Pinning in Small World Networks
}
\author{
Denis Boyer and Octavio Miramontes
}

\address{ 
Instituto de F\'\i sica, Universidad
Nacional Aut\'onoma de M\'exico, Apartado Postal 20-364, 
01000 M\'exico D.F., M\'exico
}

\date{\today}

\maketitle

\begin{abstract} 
We show that the nonequilibrium dynamics of systems
with many interacting elements located
on a small-world network can be
much slower than on regular networks. 
As an example, we study the phase ordering dynamics of the Ising model
on a Watts-Strogatz network, after a quench in
the ferromagnetic phase at zero temperature. 
In one and two dimensions, small-world features produce dynamically frozen 
configurations, disordered at large length scales, analogous of
random field models. This picture differs from the common
knowledge (supported by equilibrium results) that 
ferromagnetic short-cuts connections favor
order and uniformity. We briefly discuss some implications of these
results regarding the dynamics of social changes.
\end{abstract}

% Uncomment for twocolumn
\narrowtext
]

Small-world networks have received a great deal of attention in the past
few years, in particular for their realistic description of the topology
of the interactions that take place among populations in various biological, 
social or economical systems \cite{re:strogatz01}. 
An important feature of small worlds,
that is not shared neither by regular lattices nor random networks, 
is the interplay that exists between local (or \lq\lq physical") interactions, 
{\it e.g.} between nearest-neighbors, and non-local ones, involving
nodes (or agents) separated by large distances but 
connected through short-cuts. Among other outstanding topological properties,
the effective space dimension of such networks
grows linearly with their size \cite{re:newman99},
even if the fraction of sites with short cuts is very small.

The strong connectivity of small worlds usually 
enhances dramatically cooperative effects, as predicted by epidemic models
of spreading of diseases \cite{re:miramontes02}, or of
propagation of conventions or rumors in social systems \cite{re:zanette01}.
Naturally, many models of social dynamics have been
inspired from the Ising model \cite{re:socdyn}. 
The Ising model on a small world exhibits ferromagnetic order at low 
temperatures even in one dimension \cite{re:barrat00}, while, in higher 
dimensions, the critical temperature is increased compared with that 
of the regular lattice \cite{re:herrero02}.
In addition, the fact that the transition is of mean-field nature
agrees with the intuitive argument that each site is effectively close to
a large number of sites due to the short-cuts of the lattice. 

However, because of their inherent random topology,
one may ask whether in some situations small-world networks 
would not rather exhibit features characteristic of disordered systems. 
In this work, we 
study as a basic example the nonequilibrium dynamics
of the Ising model, as observed after a rapid quench from the 
high temperature phase to the ferromagnetic phase. 
We show that the random (all ferromagnetic) connections
that enhance ordered states at thermodynamic equilibrium,
are responsible in the present case for very slow dynamics and stabilize 
at large times configurations that, instead of being uniform,
are spatially heterogeneous. 
At zero temperature, systems do not perform long range order
dynamically, but remain asymptotically trapped in metastable states 
characterized by a finite domain size. 
These features are reminiscent of nonequilibrium processes in the random field
Ising model (RFIM) \cite{re:nattermann98}, in binary mixtures with 
fixed impurities \cite{re:jasnow01}, as well as in a few social models 
on regular lattices like the voter model \cite{re:dornic01}.
This has to be contrasted with the much more efficient phase ordering
kinetics of the Ising model on regular lattices
(or Model A in the lexicon of Hohenberg and Halperin \cite{re:hohenberg77}),
where the mean size of ordered domains grows with time as $t^{1/2}$
\cite{re:bray94}. 
Our present analysis focuses on the motion of domain walls between
\lq\lq up" and \lq\lq down" domains, and shows evidence of competing
effects between surface tension and pinning (or localizing) effects. 

We use a standard model of small-world network \cite{re:watts98} 
consisting of a regular square lattice
(or a chain in 1D) composed of $N$ nodes connected to their 
nearest neighbors. For each site, we then establish with
a probability $p$ an additional connection, or short-cut, linking the
considered site to an other site chosen at random in the lattice.
(We do not remove the nearest neighbors connections.)
For $p=0$, the lattice is regular, while for $p=1$, the network is strongly
disordered. Here, we will consider only the so-called 
\lq\lq small world" limit, that corresponds to the case $p\ll 1$, 
where connections are mainly local and only long-ranged for
a small fraction of nodes. 

On a fixed network, we then assign
to each node a spin-like variable $S_i=\pm 1$: it represents a
social convention, initially chosen at random for each node. 
At each time step, each node updates its convention in order
to reach a better consensus with the nodes it is connected to. 
In other words, the system follows a zero temperature Glauber dynamics
with the Hamiltonian $H=-\sum_{\langle i,j \rangle} J_{ij}S_i S_j$,
where the sum is performed over all possible pairs of nodes. 
$J_{ij}=1$ if sites $i$ and $j$ are connected, $J_{ij}=0$ otherwise. 
At each step, a spin is thus
chosen at random and flipped.  If $H$ decreases,
does not change, or increases, the new configuration is accepted with
probability $1$, $1/2$ and $0$, respectively.  
 
In regular networks ($p=0$), the system evolves toward
a minimum of $H$ (all $S_i$'s equal to $+1$ or $-1$).
Transient configurations
are characterized by the presence of growing and competing
ordered domains of \lq\lq up" and \lq\lq down" spins.
The large time dynamics is controlled by the motion and annihilation 
of interfaces (or domain walls) that separate these domains.
As for many other systems 
ordering in phases with broken symmetries, its time evolution
is self-similar:
The two-point correlation
function, $C(r,t)=\langle S_i(t) S_{i+r}(t)\rangle$, obeys a scaling
relation $C(r,t)=f(r/\xi(t))$, where $f$ is a scaling function, while
$\xi$, the domain size, grows as $t^{1/2}$ \cite{re:bray94}.

\begin{figure}
\vspace{-2cm}
\hspace{-1cm}\epsfig{figure=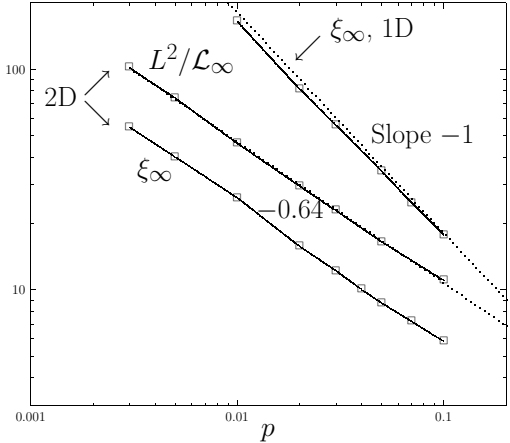,width=3.5in}
\vspace{-4cm}
\caption{Asymptotic correlation length $\xi_0$ in
lattice spacing units as a function of the 
reconnection probability $p$, in 1D and 2D. $L^2/{\cal L}_{\infty}$ 
is the interface characteristic length in 2D.}
\label{long1d2d}
\end{figure}
On the contrary, in small-world networks ($p\neq 0$, $p\ll 1$)
one observes after some time that the typical domain
size of ordered spins saturates to a finite value, which decreases when the
density of short-cuts ($p$) increases. For a one
dimensional chain (of length $L=10^5$),
we plot $\xi(t=\infty)$  as a function of $p$ in Figure \ref{long1d2d}. 
The correlation length is determined from the half-width of $C$
averaged over 10 networks and initial conditions.
A behavior $\xi(t=\infty)\propto 1/p$ can be observed. 
$1/p$ represents the characteristic size of the one dimensional
network, {\it i.e.} the average distance between two nodes
that have long range connections (or \lq\lq influent" nodes).

\begin{figure}
\vspace{0cm}
\hspace{-0.3cm}\epsfig{figure=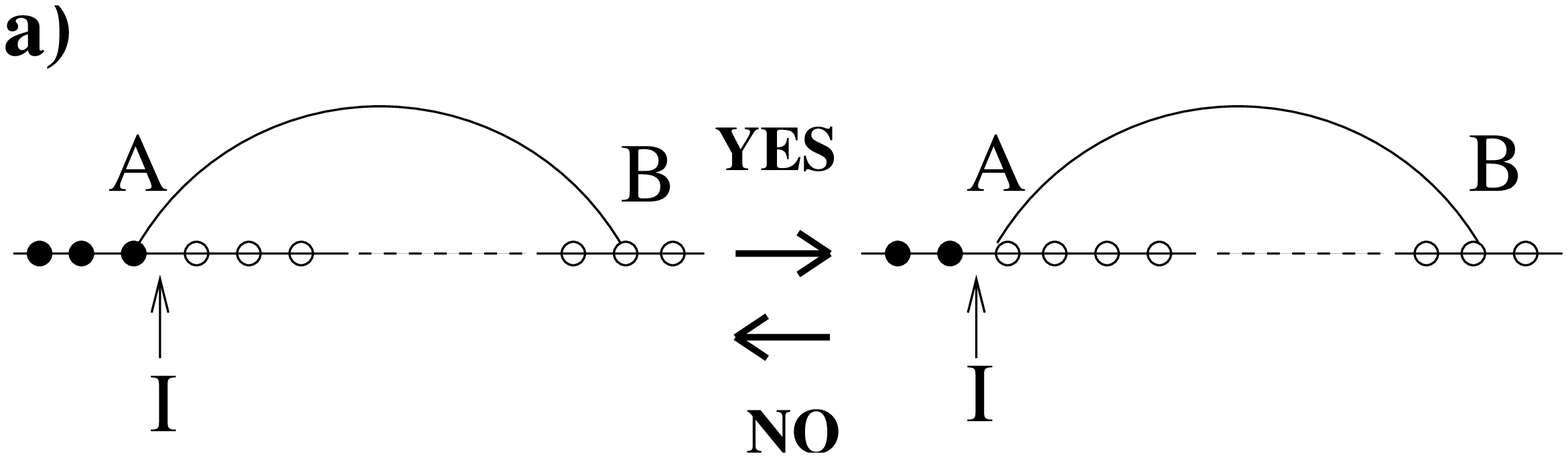,width=3.1in}
\vspace{.2cm}
\hspace{-0.3cm}\epsfig{figure=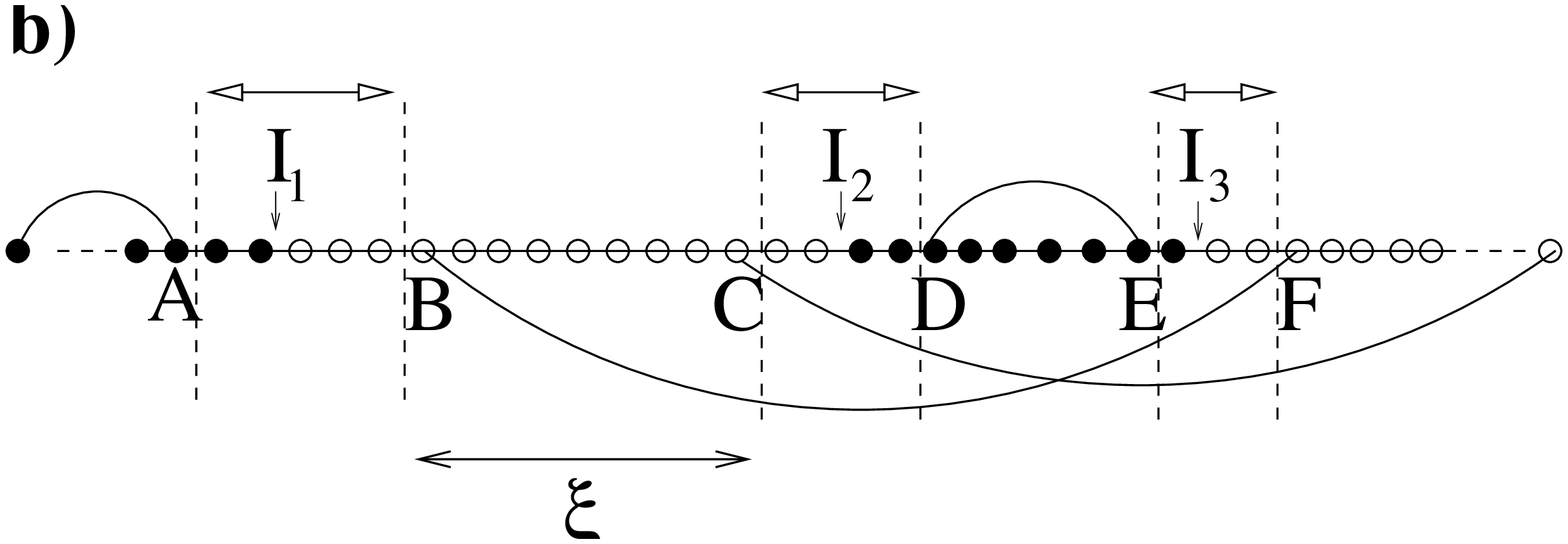,width=3.1in}
\vspace{.2cm}
\caption{Up and down domains. 
Domain walls $I_n$ become localized in 1D.}
\label{injump}
\end{figure}

Influent sites strongly affect the motion of interfaces. At low $p$, most 
of these nodes are characterized by one additional connection.
On Figure \ref{injump}$a$,
two nodes far apart, A and B, are connected, and $S_A=-S_B$.
Any interface $I$ passing through node A leftward can not jump 
back toward the right, since it is energetically unfavorable.  
Therefore, at large times, through interface motion,
influent nodes will tend to be (irreversibly) connected to nodes 
that have the same spin (a situation analogous to assortative mixing, as 
observed in some real life networks \cite{re:newman02}).

This argument can be extended to a succession of domains.
Figure \ref{injump}$b$ illustrates a typical large time configuration:
In this example, interface $I_1$ stands between two influent
nodes $A$ and $B$ with opposite spins: for 
the reason mentioned above, $I_1$ can not jump to the left of $A$, nor to
the right of $B$. The interface is then localized, {\it i.e.} restricted
to perform a random walk within the interval
[$A$,$B$]. Interface $I_1$ is therefore unable to annihilate with interface 
$I_2$, that is localized between $C$ and $D$. Hence, the two disjoint black
domains can not merge to form a bigger one, what would happen in the standard
Ising model.  The domain size,
or correlation length $\xi$, does not exceed the distance, of order 
$1/p$, that separates influent unlike (antagonist) successive nodes. 

\begin{figure}
\vspace{-2.5cm}
\hspace{-1.5cm}\epsfig{figure=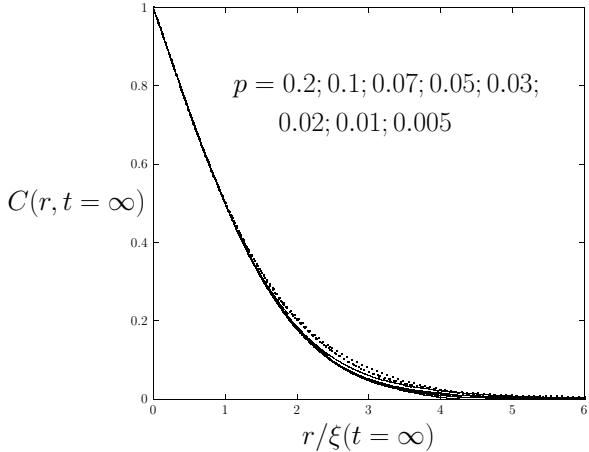,width=3.5in}
\vspace{-4.cm}
\caption{Two point correlation function in 1D (dotted lines) 
and 2D (solid lines) as a function of $r/\xi_\infty$, for various $p$.}
\label{scal1d2d}
\end{figure}

We also find that the structure of 
frozen configurations obeys a scaling relation with $p$, 
{\it i.e.} that they are statistically independent of $p$ 
via proper length rescaling. The
asymptotic correlation function $C(r,t=\infty)$ is plotted 
in Figure \ref{scal1d2d} as a function of the reduced variable 
$r/\xi_{\infty}(p)$, for various values of $p$.
Data collapse rather well on a single curve.
At short times, the kinetics is not affected by the small world 
structure of the lattice, and that $\xi(t)$ starts growing as $t^{1/2}$. 
When interfaces become localized, 
the structure can be roughly seen as the one given by the standard
Glauber dynamics of the 1D Ising model stopped at a time $(1/p)^2$. 
Since that problem obeys
dynamical scaling, frozen configuration should scale with parameter $p$.
However, this picture is not quantitatively correct, as the scaling
functions in both problems slightly differ. 

Finite, low temperature effects are quite subtle in one 
dimensional small-worlds since they do not destroy the ferromagnetic 
order observed at $T=0$, unlike in usual Ising chains.
One can interpret here the order/disorder transition temperature $T_c$ as
the temperature where ordering via interface jumps over localizing
barriers (that enable further domain merging) no longer overcomes 
disordering happening within domains (subdomain creation).
In Figure \ref{injump}$b$,
interface $I_2$ (or $I_1$) can jump in interval [$B$,$C$] at a rate
$r_a=\exp(-2/T)$. 
Besides, the rate at which any spin among  
the $p^{-1}$ spins of interval [$D$,$E$] would flip is 
$r_b=p^{-1}\exp(-4/T)$:
it is roughly the rate at which a white domain is created and can start to
grow. Qualitatively, the order/disorder transition occurs
when $r_a=r_b$. This gives: $T_c\simeq-2/\ln p$,
a expression derived (with a $\propto$ sign) in ref. \cite{re:barrat00}
using the replica method. The above relation 
may be exact as 
$p\rightarrow0$ (as the numerical prefactors in the different 
rates become irrelevant).
Simulation results (not shown) give $T_c=-2.3/\ln p$ for $p=0.03$.

\begin{figure}
\vspace{0cm}
\hspace{-0.3cm}\epsfig{figure=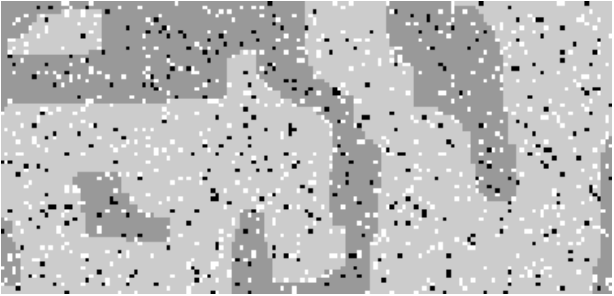,width=3.2in}
\vspace{0.5cm}
\caption{Frozen domains (in gray) for $p=0.05$ ($6\%$ of the system 
total area). The white and black dots represent the \lq\lq like" and 
\lq\lq unlike" influent nodes, respectively.}
\label{points}
\end{figure}

In two dimensions, one also observes that
random initial configurations freeze at large times. Figure \ref{long1d2d}
displays as a function of $p$ the asymptotic correlation length 
(determined from $C(r,t=\infty)$ averaged over 8 networks 
with $1500^2$ spins), as well as the length associated with interface 
density. The latter is defined as $L^2/{\cal L}$, where $L$ is the system 
linear extend and ${\cal L}$ the total length of all boundaries. 
Both length scales remain proportional to each other when varying $p$,
suggesting that frozen configurations can be characterized
by one characteristic length scale, referred to as the 
\lq\lq domain size", $R_{\infty}(p)$. Numerical results suggest that
$R_{\infty}$ varies as an inverse power-law of $p$, 
with an non-trivial exponent close to $-2/3$ over nearly two decades.
Surprisingly, $R_{\infty}$ does not scale as $p^{-1/2}$, the characteristic 
length scale of the network \cite{re:newman99}.
Once again, the spin-spin correlation function at $t=\infty$ 
scales rather well as $C(r)=f(r/\xi_\infty(p))$, see  Figure \ref{scal1d2d}.

Figure \ref{points} shows a typical frozen pattern at $p=0.05$. 
The positions of the \lq\lq influent" spins are marked by dots: in
white, those which are connected to an other influent spin 
of same sign (\lq\lq like" pairs, of number density $n_{l}$), 
in black those connected to a spin of opposite sign (\lq\lq unlike" pairs,
number density $n_u$). 
Initially, $n_l\simeq n_u$, but as coarsening proceeds, \lq\lq unlike" dots
turn more easily to \lq\lq like" than the contrary, as in
one dimension (Fig. \ref{injump}$a$). 
Once again, mixing tends to be assortative ($n_l>n_u$),
but with the increase of $n_{l}$,
at some point, there are no more possible moves toward better consensus.
We find numerically that coarsening stops and interfaces get pinned 
when $n_l\simeq 1.86\ n_u$.

The finite domain
size can be interpreted as the result of competing effects
between surface tension (the driving force for domain growth)
and energy barriers created by the multiplication of
influent \lq\lq like" sites.
We picture the
system as a collection of $L^2/R^2$ domains of radius $R$,
and estimate its energy change 
when domains coarsen from $R$ to $R+dR$ ($dR>0$).
The usual contribution from surface 
tension is $\delta E_I\propto -2L^2dR/R^2$. Meanwhile, the number of
influent nodes that flip spin is proportional 
to $2pRdR(L^2/R^2)$.  \lq\lq Like" nodes turn to \lq\lq unlike"
(with an energy cost per spin of 2), and reversely
(with an energy decrease of -2). The total energy difference thus reads
\begin{equation}
\delta E\propto \left[-\frac{2}{R^2}
+\frac{n_l-n_u}{n_l+n_u}\frac{4p}{R}\right]L^2dR.
\end{equation}
The second term is positive and dominate at large $R$. Hence, coarsening
is arrested when $\delta E=0$, or $R_{\infty}\sim p^{-1}$. 
This argument is somehow similar to the (equilibrium)
Imry-Ma argument for the RFIM \cite{re:imry75}. 
Yet, an important difference is that here the
average magnetic field felt on influent nodes (or \lq\lq impurities") is
not zero, but has been biased ($n_l\neq n_u$)
due to previous spin flips. 

\begin{figure}
\vspace{0cm}
\hspace{0.7cm}\epsfig{figure=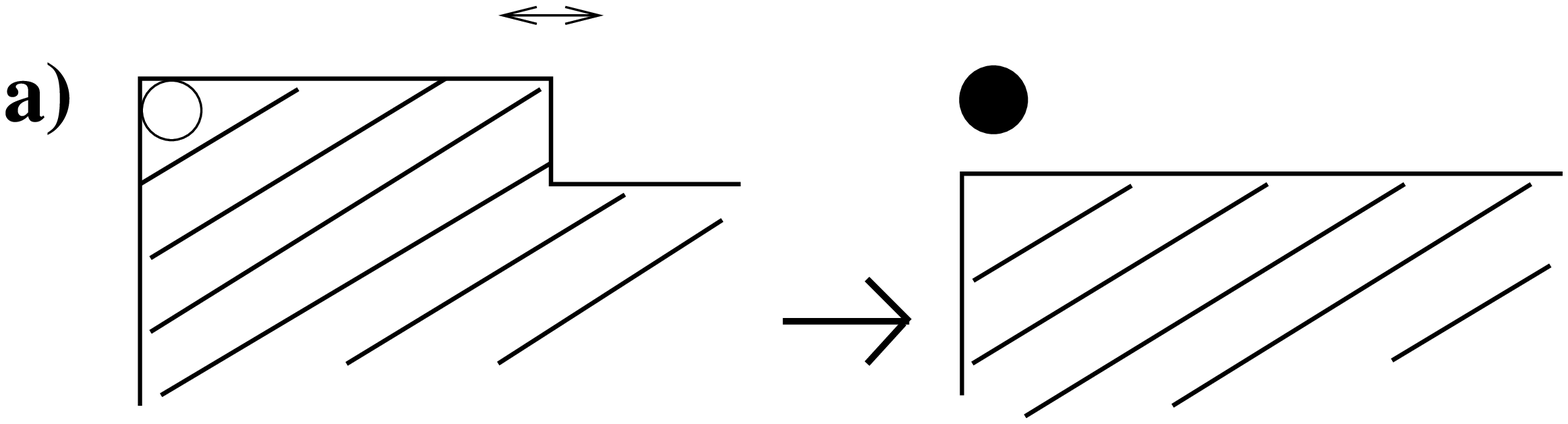,width=2.3in}

\vspace{0.3cm}
\hspace{0.7cm}\epsfig{figure=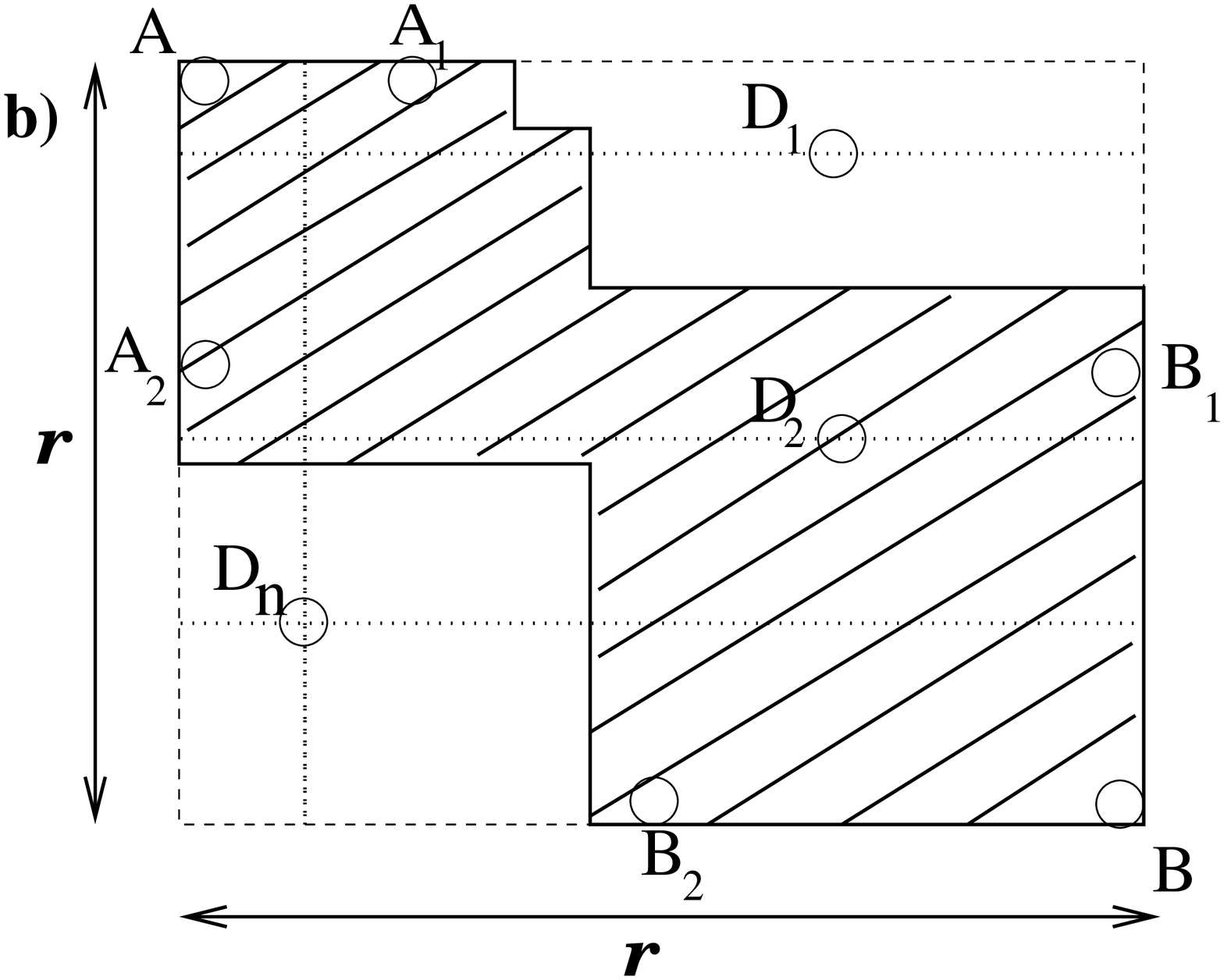,width=2.6in}
\vspace{.5cm}
\caption{
$a)$ Free and $b)$ pinned domain in presence of \lq\lq like" influent nodes.
}
\label{corners}
\end{figure}

The above continuous Imry-Ma-like argument qualitatively explains frozen 
states, but over-estimates $R_{\infty}$ ($\sim p^{-1}$ instead of $p^{-2/3}$).
The exponent $-2/3$ can be explained as an effect of the square lattice. 
As shown on Figure \ref{corners}$a$, a single influent \lq\lq like" node
located at a domain corner can disappear through the diffusive motion 
of a step.
Figure \ref{corners}$b$ represents then the simplest distribution
of \lq\lq like" nodes such that the hatched domain can not shrink. It is 
composed of two right-angle corners $\{A,A_1,A_2\}$ and $\{B,B_1,B_2\}$ 
defining a
square $r\times r$. If the other white nodes $\{D_1,..D_n\}$
contained in the square do not form any right-angle corners, then this 
region encloses 
the smallest (or \lq\lq critical") pinned domain: any bubble of hatched region 
comprised in the square and that does not contain both corners  
$\{A,A_1,A_2\}$ and $\{B,B_1,B_2\}$ will shrink. Any larger bubble will
not. We now calculate
the probability $P_{\rm freeze}(r)$ that a configuration such as represented 
in Fig. \ref{corners}$b$
has a size $r$, and then identify $r^*$ such that $P_{\rm freeze}(r^*)$ 
is maximal
with the asymptotic domain size $R_{\infty}$ in the disordered medium.

Given the node $A$ located at the origin, 
the probability that there is at least one white dot ($A_1$) on the same 
line within a distance $r$ is $P_1(r)=1-(1-p_l)^r$, with $p_l/p= 
n_l/(n_l+n_u)$ the fraction of influent nodes that are \lq\lq like" 
(here, the numerical value of this ratio $-$ close to 0.65 $-$ 
is unimportant and could be set to 1).
Therefore, 
$P_{\rm freeze}(r)=p_l[P_1(r)]^4 P_2(r)$,
with $P_2$ the probability that 
the $D_n$'s do not form right-angle corners, {\it i.e.} 
that each node $D_i$ is at least located on a line or a column not occupied 
by an other $D_j$ (see the dotted lines in Fig. \ref{corners}b). 
$P_2$ can be approximated as
\begin{equation}\label{P2}
P_2(r)\simeq\sum_{n=0}^{(r-2)^2}(1-p_l)^{(r-2)^2-n}p_l^n 
C^n_{(r-2)^2}
[1-P_1(r)]^n, 
\end{equation}
or $P_2(r)\simeq (1-p_lP_1(r))^{(r-2)^2}$. 
In the sum (\ref{P2}) we have multiplied the probability of having 
$n$ white dots inside the square by the probability $[1-P_1(r)]^n$ that
$n$ independent dots have no neighbors on the same line within $r$.
For $n$ small, no or few corners can be formed anyway, so that relation 
(\ref{P2}) slightly under-estimated $P_2$, since a 
small fraction of empty sites are counted twice ($[1-P_1(r)]^n\lesssim 1$).  
For $n$ large, on the contrary, relation (\ref{P2}) over-estimates $P_2$, 
since it is
impossible to locate many dots without forming corners
(while $[1-P_1(r)]^n$ is small but $\neq 0$). We suppose that both errors 
compensate.
This factorization enable us to compute the most probable
square size $r^*$ analytically.
$P_1(r)$ increases with $r$, $P_2(r)$ decreases with $r$, and 
$P_{\rm freeze}(r)$ has one single maximum.
Assuming $p\ll 1$, $r\gg 1$, $rp\ll 1$,  we find that
$r^*=p_l^{-2/3}\propto p^{-2/3}$, in agreement with the numerical results. 

To summarize, we have shown on an example that
assortative mixing in small world networks can dynamically generate 
frozen metastable states : At large times, some influent nodes have simply
no immediate interest to evolve. 
These results suggest that long term dynamics in highly connected
social systems can produce spatial heterogeneities 
(or segregation), despite that these configurations are not
the most desired ones by individual agents. A similar picture,
in agreement with some empirical observations,
was drawn recently from antiferromagnetic models on scale-free networks 
\cite{re:stauffer}.
Right after strong political changes (in Eastern European countries
in 1989, in Mexico in 2000) the evolution of reforms can be fast, but 
rapidly social inertia takes over and renders further adjustments 
difficult or null.
Physically speaking, the response of social systems to external forcings
({\it i.e.} large-scale policies) is susceptible
to exhibit some of the interesting features known for disordered systems
\cite{re:nattermann98}.

While revising the manuscript, we became aware of a similar study
on the voter model on small-worlds \cite{re:castellano02}.
We acknowledge fruitful discussions with G. Cocho, J. Vi\~nals and R. Boyer.

\bibliographystyle{prsty}

\end{document}